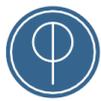

Research Article

# Effects of kappa-opioid agonist U-50488 and p38 MAPK inhibitor SB203580 on the spike activity of pyramidal neurons in the basolateral amygdala

Konstantin Y. Kalitin[1,2], Alexander A. Spasov[1,2], Olga Y. Mukha[1]

1 Volgograd State Medical University, 1 Pavshikh Bortsov Sq., Volgograd 400131 Russia
2 Volgograd Medical Research Center, 1 Pavshikh Bortsov Sq., Volgograd 400131 Russia

Corresponding author: Olga Y. Mukha (olay.myha14@gmail.com)

## Abstract

**Introduction**: Kappa-opioid receptor (KOR) signaling in the basolateral amygdala (BLA) underlies KOR agonist-induced aversion. In this study, we aimed to understand the individual and combined effects of KOR agonist U-50488 and p38 MAPK inhibitor SB203580 on the spiking activity of pyramidal neurons in the BLA to shed light on the complex interplay between KORs, the p38 MAPK, and neuronal excitability. .

**Materials and methods:** Electrophysiological experiments were performed using the patch-clamp technique in the whole-cell configuration. Rat brain slices containing the amygdala were prepared, and pyramidal neurons within the BLA were visually patched and recorded in the current clamp mode. The neurons were identified by their accommodation properties and neural activity signals were amplified and analyzed. Using local perfusion, we obtained three dose-response curves for: (a) U-50488 (0.001–10 μM); (b) U-50488 (0.001–10 μM) in the presence of SB203580 (1 μM); and (c) U-50488 (0.01–10 μM) in the presence of SB203580 (5 μM).

**Results:** After the application of U-50488, pyramidal neurons had a higher action potential firing rate in response to a current injection than control neurons (p<0.001). The dose-dependent curves we obtained indicate that the combination of U-50488 and SB203580 results in non-competitive antagonism. This conclusion is supported by the observed change in the curve's slope with reduction in the maximum effect of U-50488. Thus, it can be assumed that the increase in spike activity of pyramidal neurons of the amygdala is mediated through the beta-arrestin pathway. When this pathway is blocked, the spike activity reverts to its baseline level.

**Conclusion:** Our study found that the KOR agonist-induced spiking activity of the BLA pyramidal neurons is mediated by the beta-arrestin pathway and can be suppressed by the application of the p38 MAPK inhibitor SB203580.

## Graphical Abstract

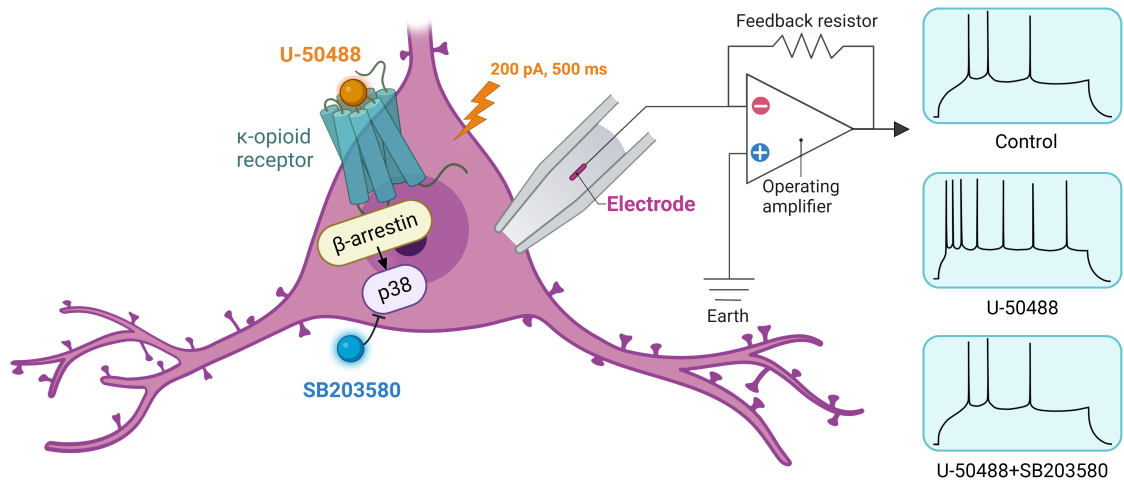

Whole-cell patch-clamp recordings from basolateral amygdala pyramidal cells showing spiking patterns in response to 200 pA, 500 ms-long current injections following the treatment with the kappa-opioid agonist U-50488 and the p38 MAPK inhibitor SB203580. U-50488 increased the excitability of pyramidal neurons, an effect that was blocked by SB203580, suggesting the involvement of the beta-arrestin pathway in basolateral amygdala neural activity.

## Keywords

kappa-opioid receptor, p38 MAPK, amygdala, pyramidal neurons, patch clamp, electrophysiology, aversion

## Introduction

The amygdala is instrumental in processing emotions associated with aversion (Michely et al. 2020). It plays a role in fear conditioning, interpretation of affective stimuli, regulation of stress responses, and modulation of pain perception. Moreover, its extensive interconnections with other brain regions allow the amygdala to integrate sensory inputs into emotional contexts. The basolateral amygdala (BLA), a cluster of nuclei within the amygdala, is pivotal in the formation and retrieval of emotionally charged memories and has a significant influence on emotional behavior. The BLA is a central component of the anxiety circuit, processing executive and sensory information, and relaying it to brain areas that evoke physical and psychological anxiety responses. Notably, the activity of the BLA's excitatory glutamatergic pyramidal neurons is closely associated with anxiety behavior and other aversive effects (Wang et al. 2011; Janak and Tye 2015).

Kappa-opioid receptors (KORs) represent a class of G-protein-coupled receptors that are widely distributed throughout the central nervous system and extensively expressed in limbic brain areas including the basolateral amygdala, central nucleus of amygdala, extended amygdala, hypothalamus, and hippocampus. The activation of KORs by agonists like U-50488 leads to a series of physiological and psychological effects. These effects encompass analgesia, increased diuresis, and antipruritic activity (Khan et al. 2022). KORs are involved in various behavioral aversive effects such as depression, anhedonia, dysphoria, and anxiety (Knoll et al. 2011).

It was shown that KOR signaling in the BLA regulates conditioned fear and anxiety in rats (Knoll et al. 2011). Anxiolytic effects have been observed following the systemic administration of KOR antagonists, as evidenced in the elevated plus maze test. The pretreatment with a specific KOR antagonist, nor-binaltorphimine (nor-BNI), has been found to prevent the conditioned place preference to intravenous gabapentin in the spinal nerve ligation model of neuropathic pain. This outcome suggests that nor-BNI effectively mitigates the aversiveness of ongoing pain. Recent studies suggest that KORs play a significant role in modulating neuronal excitability. Nor-BNI reduced the synaptically evoked spiking of amygdala neurons in brain slices from rats with spinal nerve ligation (Navratilova et al. 2019).

The mitogen-activated protein kinase (MAPK) pathway, specifically the p38 MAPK, is involved in several cellular processes, from inflammation to cell differentiation and death. In the context of aversion, p38 MAPK in the amygdala has been associated with dysphoria. Kappa-opioid receptor agonist U-50488 induced significant place aversion in mice, as measured by the place conditioning paradigm, accompanied by significant activation of p38 MAPK in the amygdala, but not in the hippocampus and nucleus accumbens. Intra-amygdalar microinjection of the specific p38 MAPK inhibitor SB203580 completely blocked U-50488-induced conditioned place aversion in mice (Zan et al. 2016). The findings indicate that the activation of p38 MAPK in the amygdala is essential for mediating the aversive behavior induced by KOR agonists.

In this study, we aimed to understand the individual and combined effects of U-50488 and SB203580 on the spiking activity of pyramidal neurons in the BLA (region critical for aversion-related behavior) to shed light on the complex interplay between KORs, the p38 MAPK, and neuronal excitability.

# Materials and methods

### Experimental animals

Adult male Sprague Dawley rats, weighing 230–250 g, were used in these studies. Animals were individually housed, maintained in a 12 h light/12 h dark cycle environment with controlled temperature (22±2 °C), and provided food and water ad libitum. All procedures complied with the Principles of Good Laboratory Practice (Interstate Standard of the Russian Federation GOST 33647-2015) and the Order of the Ministry of Health and Social Development of the Russian Federation dated 01.04.2016 No. 199n "On Approval of the Rules of Laboratory Practice". Experiment protocols were approved by the Regional Research Ethics Committee of Volgograd Region (registration number IRB00005839 IORG0004900, Minutes No. 2022/096 of 21.01.2022).

### Drugs

The following drugs were used: U-50488 (Sigma-Aldrich) and SB203580 (Sigma-Aldrich).

### Slice preparation

Brain slices were harvested from Sprague Dawley rats weighing 240–270 g. Coronal brain slices, 500 μm thick and containing the BLA, were cut 2.5–3.5 mm caudal to the bregma using a vibratome (Campden 7000smz-2, UK). Each slice was transferred to a recording chamber, submerged in artificial cerebrospinal fluid (aCSF) maintained at 31±1°C with a superfusion rate of 2 mL/min. The aCSF was composed of 117 mM NaCl, 4.7 mM KCl, 1.2 mM $NaH_2PO_4$, 2.5 mM $CaCl_2$, 1.2 mM $MgCl_2$, 25 mM $NaHCO_3$, and 11 mM glucose, continuously aerated with a mixture of 95% $O_2$ and 5% $CO_2$ at pH 7.4. We used only one or two slices per animal and recorded one neuron per slice.

### Patch-clamp recording

The neurons were visually patched under an X40 water immersion objective of the microscope (Olympus BX51, Japan) in the whole-cell configuration and were recorded in the current clamp mode. Pyramidal cells in the BLA were identified based on their accommodation properties in response to a sustained depolarizing intracellular current injection. The recording pipettes made from borosilicate glass were filled with a solution containing: 122 mM K-gluconate, 5 mM NaCl, 0.3 mM $CaCl_2$, 2 mM $MgCl_2$, 1 mM EGTA, 10 mM HEPES, 5 mM $Na_2$-ATP, and 0.4 mM $Na_2$-GTP, pH adjusted to 7.2–7.3 with KOH (osmolarity adjusted to 280 mOsm/kg with sucrose). The recordings were amplified by a HEKA Patch Clamp EPC10 USB (HEKA Elektronik, USA) and analyzed by PatchMaster (HEKA Elektronik, USA) software.

We obtained three dose-response curves to evaluate the effects of U-50488, a kappa opioid receptor agonist, on the spike activity of pyramidal neurons both in the absence and presence of SB203580, a specific inhibitor of p38 MAP kinase. The three experimental conditions included: (a) treatment with U-50488, applied by local perfusion at concentrations ranging from 0.001 to 10 μM; (b) treatment with a combination of U-50488 (0.001–10 μM) and SB203580 (1 μM); and (c) treatment with a combination of U-50488 (0.01–10 μM) and SB203580 (5 μM). aCSF served as vehicle control in all experiments.

### Statistical analysis

Statistical analysis was conducted with two-way ANOVA followed by Dunnet *post hoc* test using GraphPad Prism 9.5. Statistical significance was accepted at the level $p<0.05$.

# Results

BLA neurons come in two types: pyramidal neurons and interneurons. The neurons were distinguished based on their firing properties (Sah et al. 2003). A 500-ms intracellular current injection through the recording microelectrode into pyramidal neurons (the main cell type found in the BLA) evoked action potentials exhibiting spike frequency adaptation due to activation of a slow afterhyperpolarization. In contrast, similar current injections into interneurons resulted in a high-frequency series of action potentials without frequency adaptation.

To evaluate SB203580's impact on U-50488-induced spike activity in amygdala pyramidal neurons, we co-treated brain slices with p38 MAPK inhibitor SB203580 and various concentrations of U-50488 and measured neuronal spike activity over 500 ms (Fig. 2). The application of SB203580 in the absence of U-50488 did not lead to a statistically significant change in spiking frequency, suggesting that KOR activation may be a prerequisite for the observed SB203580 action.

Pyramidal neurons after U-50488 application had a higher action potential firing rate in response to a current injection than control neurons ($p<0.001$), which was consistent with the hypothesis that U-50488 enhances neuronal excitability (Figs 1 and 2). At concentrations exceeding 1 μM, U-50488 induced a plateau in spike frequency. The dose-response curve closely fit the experimental data, with a coefficient of determination ($R^2$) of 0.96, indicating a strong correlation between U-50488 concentration and neuronal spike activity.

When U-50488 was combined with 1 μM of SB203580, the maximal effect ($E_{max}$) decreased to 70.44%, indicating that SB203580 had an inhibitory effect on the U-50488 induced spike activity of pyramidal neurons ($p<0.05$). The $R^2$ value of 0.93 still indicated a good fit of the data. The decreased slope, compared to the top curve, implied that the increase in spike activity was less steep in the presence of SB203580.

With an increased concentration of SB203580 at 5 μM, there was a more pronounced decrease in the $E_{max}$ to 25.73%. The $R^2$ value decreased to 0.79. The slope also decreased, reflecting a diminished responsiveness of the neurons to increasing concentrations of U-50488 in the presence of 5 μM SB203580.

Thus, the p38 MAP kinase pathway, which was inhibited by SB203580, appears to play a significant role in the mechanism through which U-50488 increases the spike activity of pyramidal neurons.

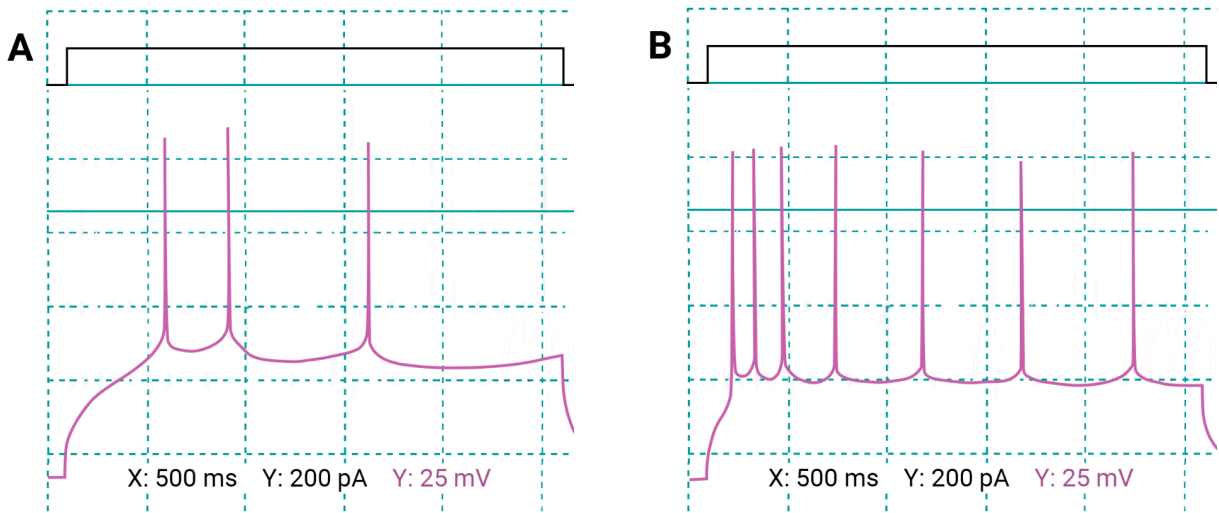

**Figure 1.** Representative traces of action potentials generated by intracellular depolarizing current injection (200 pA, 500 ms) in BLA pyramidal neurons after (A) aCSF and (B) U-50488 (1 μM) application.

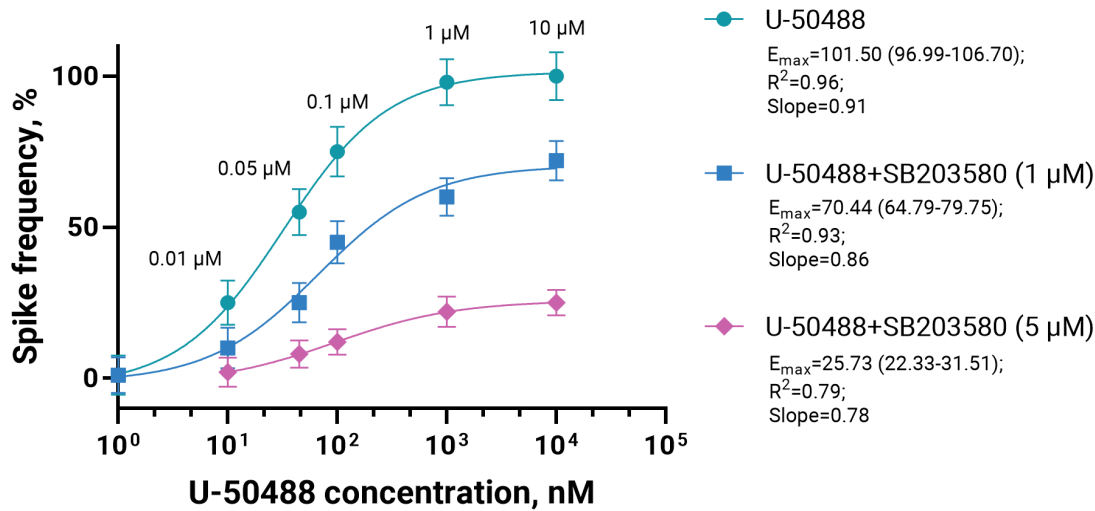

**Figure 2.** Dose-dependent effects of U-50488 (0.001–10 μM) alone and in combination with SB203580 (1 and 5 μM) on the spike activity of the BLA pyramidal neurons in response to current injection, presented as mean±SEM of the number of spikes recorded in 500 ms intervals. The maximal effect

# Discussion

The amygdala, particularly the basolateral amygdala, plays a crucial role in processing negative emotions like fear and anxiety, making it relevant to the experience of aversion. On the other hand, the activation of kappa opioid receptors correlates with aversive outcomes, such as prodepressant and anxiogenic responses, and facilitates stress-induced reinstatement. It is consistent with the observation that acute stress-induced anxiety is mediated by the release of dynorphin, an endogenous ligand for KOR, in the BLA (Bruchas et al. 2009; Varlinskaya et al. 2020). The administration of norBNI, a KOR antagonist, into the BLA blocks this effect. This implies a direct role of KOR activation in stress and anxiety, and suggests that antagonizing KOR can mitigate anxiety-like behaviors. Knoll and colleagues found that the injection of kappa-opioid receptor antagonist into the BLA produced an anxiolytic-like response in the elevated plus-maze test (Knoll et al. 2011).

The BLA's efferent projections extend to various brain regions, enabling it to modulate diverse aspects of cognitive and emotional functions. One of the principal projection targets of the BLA is the prefrontal cortex (PFC). This connection facilitates the integration of emotional and cognitive processes (Ji et al. 2010). In addition, it was found that pain can lead to BLA-dependent impairment of mPFC function, resulting in a deficit in cognitive decision-making (difficulties in making decisions that involve complex cognitive processing) (Ji et al. 2010). The communication between the BLA and mPFC is complex and is still a topic of ongoing research

Moreover, the BLA indirectly interacts with the brainstem and hypothalamus, regions that are involved in the autonomic and physiological responses to stress and aversive stimuli (Lamotte et al. 2021). These connections allow the BLA to modulate physiological reactions, such as changes in heart rate, respiration, and hormonal responses, which are typically associated with aversion (Dampney 2019). The BLA, through these pathways, can influence the body's stress response mechanisms to prepare the individual to confront or avoid the aversive stimuli.

Wang et al. (2011) found that in freely moving mice, a group of neurons in the BLA fires tonically under anxiety conditions during both open-field and elevated plus-maze tests (Wang et al. 2011). The firing patterns of these neurons exhibited a slow onset and progressively increased firing rates, characteristics more typically associated with pyramidal neurons. Notably, a strong correlation was observed between the excitability of these BLA neurons and the anxiety levels of the mice. Subsequent research established that the dynorphin and kappa-opioid system in the basolateral amygdala mediates anxiety-like behavior (Bruchas et al. 2009). Pyramidal neurons are the main output cells of the BLA, and it is believed that kappa opioid agonists primarily affect these neurons.

It has also been established that beta-arrestin/p38 MAPK post-receptor signaling plays a crucial role in mediating the aversive effects of KOR agonists. The aversive effect of U-50488 was demonstrated to be completely eliminated when the p38 MAPK inhibitor SB203580 was administered directly into the amygdala (Zan et al. 2016).

Our findings demonstrate that the beta-arrestin/p38 pathway mediates an increase in spike activity of BLA pyramidal neurons. Blocking this pathway causes the spike activity to revert to its baseline level. The dose-dependent curves we obtained indicate that the combination of U-50488 and SB203580 results in non-competitive antagonism. This conclusion is supported by the observed change in the curve's slope with reduction in the maximum effect of U-50488.

The question remains whether an increase in spike firing frequency of BLA pyramidal neurons results directly from the activation of kappa-opioid receptors located on these neurons. Alternatively, kappa-opioid agonists may target other neurons that secondarily influence the excitability of BLA pyramidal neurons. Selective KOR inhibitor Nor-BNI has been observed to reduce the frequency of spontaneous inhibitory synaptic currents in the central amygdala pyramidal neurons, without affecting their amplitude, suggesting a presynaptic mode of action (Yakhnitsa et al. 2022).

The findings suggest that the activation of pyramidal neurons in the BLA following the administration of kappa opioid agonists is potentially associated with the inhibition of GABA interneuron activity. Glutamatergic transmission in the BLA may also play a significant role. However, studies have shown that the KOR agonist U69593 does not induce changes in either GABAergic or glutamatergic transmission in the adult BLA (Przybysz et al. 2017).

Our study found that the KOR agonist-induced spiking activity of pyramidal neurons is mediated by the beta-arrestin pathway and can be suppressed by the application of the p38 MAPK inhibitor SB203580. The obtained results improve our understanding of the role that kappa opioid receptors play in regulating the amygdala and associated brain areas. This knowledge is crucial for elucidating the aversive effects and cognitive dysfunction induced by kappa opioid agonists. The effects of various kappa opioid ligands, including biased agonists, on the spiking activity of BLA pyramidal neurons have yet to be thoroughly assessed.

## Conclusion

The paper provides insights into the complex interactions between pyramidal neurons of the BLA, kappa opioid receptors, MAPK p38, and associated pharmacological agents. It was shown that the kappa opioid agonist U-50488 enhances the excitability of pyramidal neurons, an effect that can be modulated by the MAPK p38 inhibitor SB203580, suggesting the involvement of the beta-arrestin pathway in BLA neural activity. A comprehensive evaluation of the impact of kappa opioid ligands on the spiking activity of BLA pyramidal neurons is crucial for understanding the cognitive and aversive effects induced by kappa opioid agonists.

**Conflict of interest**

The authors have declared that no competing interests exist.

**Funding**

The authors have no funding to report.

**Data availability**

All of the data that support the findings of this study are available in the main text.

## References

- Bruchas MR, Land BB, Lemos JC, Chavkin C (2009) CRF1-R activation of the dynorphin/kappa opioid system in the mouse basolateral amygdala mediates anxiety-like behavior. PloS One 4(12): e8528. https://doi.org/10.1371/journal.pone.0008528 [PubMed] [PMC]
- Dampney RAL (2019) Chapter 28 – Central Mechanisms Generating Cardiovascular and Respiratory Responses to Emotional Stress. In: Fink G (Eds) Stress: Physiology, Biochemistry, and Pathology, Academic Press, pp. 396–397. https://doi.org/10.1016/B978-0-12-813146-6.00028-X
- Franco-García A, Fernández-Gómez FJ, Gómez-Murcia V, Hidalgo JM, Milanés MV, Núñez C (2022) Molecular mechanisms underlying the retrieval and extinction of morphine withdrawal-associated memories in the basolateral amygdala and dentate gyrus. Biomedicines 10(3): 588. https://doi.org/10.3390/biomedicines10030588 [PubMed] [PMC]
- Franco-García A, Gómez-Murcia V, Fernández-Gómez FJ, González-Andreu R, Hidalgo JM, Victoria Milanés M, Núñez C (2023) Morphine-withdrawal aversive memories and their extinction modulate H4K5 acetylation and Brd4 activation in the rat hippocampus and basolateral amygdala. Biomedicine & Pharmacotherapy 165: 115055. https://doi.org/10.1016/j.biopha.2023.115055 [PubMed]
- Janak PH, Tye KM (2015) From circuits to behaviour in the

amygdala. Nature 517(7534): 284–292. https://doi.org/10.1038/nature14188 [PubMed] [PMC]

Ji G, Sun H, Fu Y, Li Z, Pais-Vieira M, Galhardo V, Neugebauer V (2010) Cognitive impairment in pain through amygdala-driven prefrontal cortical deactivation. The Journal of Neuroscience 30(15): 5451–5464. https://doi.org/10.1523/JNEUROSCI.0225-10.2010 [PubMed] [PMC]

Khan MI H, Sawyer BJ, Akins NS, Le HV (2022) A systematic review on the kappa opioid receptor and its ligands: New directions for the treatment of pain, anxiety, depression, and drug abuse. European Journal of Medicinal Chemistry 243: 114785. https://doi.org/10.1016/j.ejmech.2022.114785 [PubMed]

Knoll AT, Muschamp JW, Sillivan SE, Ferguson D, Dietz DM, Meloni EG, Carroll FI, Nestler EJ, Konradi C, Carlezon WA Jr (2011) Kappa opioid receptor signaling in the basolateral amygdala regulates conditioned fear and anxiety in rats. Biological Psychiatry 70(5): 425–433. https://doi.org/10.1016/j.biopsych.2011.03.017 [PubMed] [PMC]

Lamotte G, Shouman K, Benarroch EE (2021) Stress and central autonomic network. Autonomic Neuroscience: Basic & Clinical 235: 102870. https://doi.org/10.1016/j.autneu.2021.102870 [PubMed]

Michely J, Rigoli F, Rutledge RB, Hauser TU, Dolan RJ (2020) Distinct Processing of Aversive Experience in Amygdala Subregions. Biological psychiatry. Cognitive Neuroscience and Neuroimaging 5(3): 291–300. https://doi.org/10.1016/j.bpsc.2019.07.008 [PubMed] [PMC]

Navratilova E, Ji G, Phelps C, Qu C, Hein M, Yakhnitsa V, Neugebauer V, Porreca F (2019) Kappa opioid signaling in the central nucleus of the amygdala promotes disinhibition and aversiveness of chronic neuropathic pain. Pain 160(4): 824–832. https://doi.org/10.1097/j.pain.0000000000001458 [PubMed] [PMC]

Przybysz KR, Werner DF, Diaz MR (2017) Age-dependent regulation of GABA transmission by kappa opioid receptors in the basolateral amygdala of Sprague-Dawley rats. Neuropharmacology 117: 124–133. https://doi.org/10.1016/j.neuropharm.2017.01.036 [PubMed] [PMC]

Sah P, Faber ES, Lopez De Armentia M, Power J (2003) The amygdaloid complex: anatomy and physiology. Physiological Reviews 83(3): 803–834. https://doi.org/10.1152/physrev.00002.2003 [PubMed]

Varlinskaya EI, Johnson JM, Przybysz KR, Deak T, Diaz MR (2020) Adolescent forced swim stress increases social anxiety-like behaviors and alters kappa opioid receptor function in the basolateral amygdala of male rats. Progress in Neuro-psychopharmacology & Biological Psychiatry 98: 109812. https://doi.org/10.1016/j.pnpbp.2019.109812 [PubMed] [PMC]

Wang DV, Wang F, Liu J, Zhang L, Wang Z, Lin L (2011) Neurons in the amygdala with response-selectivity for anxiety in two ethologically based tests. PloS One 6(4): e18739. https://doi.org/10.1371/journal.pone.0018739 [PubMed] [PMC]

Yakhnitsa V, Ji G, Hein M, Presto P, Griffin Z, Ponomareva O, Navratilova E, Porreca F, Neugebauer V (2022) Kappa opioid receptor blockade in the amygdala mitigates pain like-behaviors by inhibiting corticotropin releasing factor neurons in a rat model of functional pain. Frontiers in Pharmacology 13: 903978. https://doi.org/10.3389/fphar.2022.903978 [PubMed] [PMC]

Zan GY, Wang Q, Wang YJ, Chen JC, WuX, Yang CH, Chai JR, Li M, Liu Y, Hu XW, Shu XH, Liu JG (2016) p38 mitogen-activated protein kinase activation in amygdala mediates κ opioid receptor agonist U50,488H-induced conditioned place aversion. Neuroscience 320: 122–128. https://doi.org/10.1016/j.neuroscience.2016.01.052 [PubMed]

# Author contributions

- **Konstantin Y. Kalitin**, PhD in Medicine, Associate Professor of the Department of Pharmacology and Bioinformatics, Volgograd State Medical University, Volgograd, Russia; Senior Researcher, Laboratory of Metabotropic Drugs, Scientific Center for Innovative Drugs, Volgograd State Medical University, Volgograd, Russia; Researcher, Laboratory of Experimental Pharmacology, Volgograd Medical Research Center, Volgograd, Russia; e-mail: kkonst8@yandex.ru; **ORCID ID** https://orcid.org/0000-0002-0079-853X. Conceptualization, setting primary objectives, carrying out experimental procedures, literature review, analyzing and interpreting the data, writing the initial draft of the article, contributing to the overall structure and design, finalizing the article for submission, designing visual data representations, applying statistical, mathematical, and computational methods for data analysis.

- **Alexander A. Spasov**, PhD in Medicine, Professor, memberof the Russian Academy of Sciences; Head of the Department of Pharmacology and Bioinformatics, Volgograd State Medical University, Volgograd, Russia; Head of the Laboratory of Experimental Pharmacology, Volgograd Medical Research Center, Volgograd, Russia; Head of the Department of Pharmacology and Bioinformatics, Scientific Center for Innovative Drugs, Volgograd State Medical University, Volgograd, Russia; e-mail: aaspasov@volgmed.ru; **ORCID ID** https://orcid.org/0000-0002-7185-4826. Concept statement, supervision and mentorship, workflow control, critical revision with valuable intellectual investment, contributing to the overall structure and design, ensuring the integrity and consistency of the entire article.

- **Olga Y. Mukha**, PhD candidate, postsecondary teacher of the Department of Pharmacology and Bioinformatics, Volgograd State Medical University, Volgograd, Russia; e-mail: olay.myha14@gmail.com; **ORCID ID** https://orcid.org/0000-0002-0429-905X. Task management, experimental work, literature review, data collection, drafting the manuscript, contributing to the scientific layout, ensuring the integrity and consistency of the entire article, including its final version, application of statistical, mathematical, and computational methods for data analysis, preparation of the final manuscript, interaction with the editors and reviewers.